\documentstyle[12pt]{article}

\begin{document} 

\title{Dynamics in Non-Globally-Hyperbolic Static Spacetimes II:
General Analysis of Prescriptions for Dynamics} 

\author{Akihiro Ishibashi$^{\dag}$ and Robert M. Wald$^{\ddag}$ \\ \\ 
 {\it $^{\dag}$Department of Applied Mathematics and Theoretical Physics,} \\ 
 {\it Centre for Mathematical Sciences,} \\
 {\it University of Cambridge, Wilberforce Road,}\\ 
 {\it Cambridge CB3 0WA, UK} \\ \\
 {\it $^{\ddag}$Enrico Fermi Institute and Department of Physics}\\ 
 {\it University of Chicago}\\ 
 {\it 5640 S. Ellis Avenue}\\ 
 {\it Chicago, Illinois 60637-1433, USA} 
         } 

\maketitle

\begin{abstract}

It was previously shown by one of us that in any static,
non-globally-hyperbolic, spacetime it is always possible to define a
sensible dynamics for a Klein-Gordon scalar field. The prescription
proposed for doing so involved viewing the spatial derivative part,
$A$, of the wave operator as an operator on a certain $L^2$ Hilbert
space $\mathcal H$ and then defining a positive, self-adjoint operator
on $\mathcal H$ by taking the Friedrichs extension (or other positive
extension) of $A$. However, this analysis left open the possibility
that there could be other inequivalent prescriptions of a completely
different nature that might also yield satisfactory definitions of the
dynamics of a scalar field. We show here that this is not the
case. Specifically, we show that if the dynamics agrees locally with
the dynamics defined by the wave equation, if it admits a suitable
conserved energy, and if it satisfies certain other specified
conditions, then it must correspond to the dynamics defined by
choosing some positive, self-adjoint extension of $A$ on $\mathcal
H$. Thus, subject to our requirements, the previously given
prescription is the only possible way of defining the dynamics of a
scalar field in a static, non-globally-hyperbolic, spacetime. In a
subsequent paper, this result will be applied to the analysis of
scalar, electromagnetic, and gravitational perturbations of
anti-de Sitter spacetime. By doing so, we will determine all possible
choices of boundary conditions at infinity in anti-de Sitter spacetime
that give rise to sensible dynamics.

\end{abstract}

\section{Introduction}

Let $(M, g_{ab})$ be a spacetime and let $\Phi$ be a real\footnote{The
analysis of the dynamics of a complex scalar field follows immediately
from that of the real scalar field by treating the real and imaginary
parts separately.} scalar field in this spacetime that satisfies the
Klein-Gordon equation
\begin{equation}
\nabla^a \nabla_a \Phi - m^2 \Phi = 0 \,. 
\label{kg}
\end{equation}
It is well known (see, e.g., \cite{w}) that if $(M, g_{ab})$ is
globally hyperbolic and if $\Sigma$ is a Cauchy surface for $(M,
g_{ab})$, then for each $(\phi_0, \dot{\phi}_0) \in C^\infty(\Sigma)
\times C^\infty(\Sigma)$, there exists a unique $\Phi \in C^\infty(M)$
that satisfies eq.~(\ref{kg}) and is such that $\Phi|_\Sigma = \phi_0$
and $t^a \nabla_a \Phi|_\Sigma = \dot{\phi}_0$, where $t^a$ is a given
(not necessarily unit) normal to $\Sigma$.

However, no similar result holds, in general, when $(M, g_{ab})$ fails
to be globally hyperbolic. Given suitable smooth data for $\Phi$ at
some ``time'' (i.e., on a spacelike slice, $\Sigma$), the partial
differential equation (\ref{kg}) need not, in general, admit any
solutions corresponding to this data and, even when a solution exists,
it need not be unique. Given the infinite variety of singularities and
causal pathologies that can occur in an arbitrary
non-globally-hyperbolic spacetime, it is far from clear how to even
attempt to give a unique prescription for defining dynamics.
Nevertheless, it is of interest to analyze dynamics in
non-globally-hyperbolic spacetimes and determine whether there exist
prescriptions that give rise to acceptable, unique dynamics, since
this might give hints as to whether and how the spacetime
singularities themselves might be ``resolved'' in quantum gravity.  It
is well known that classical general relativity predicts the
occurrence of singularities. If these singularities are ``resolved''
by quantum gravity, then one might expect to have well defined,
deterministic predictions in situations corresponding to the presence
of classical singularities.

As a first step in the direction of analyzing dynamics in
non-globally-hyperbolic spacetimes, attention was restricted in
\cite{w1} to static, non-globally-hyperbolic spacetimes that possess a
hypersurface $\Sigma$ orthogonal to the static Killing field, $t^a$,
such that the orbits of $t^a$ are complete and each orbit intersects
$\Sigma$ once and only once. No causal pathologies occur in such
spacetimes (since they are automatically stably causal), but there is
still an infinite variety of possible singular behaviour. It was shown
in \cite{w1} that in such spacetimes, a sensible dynamics always can
be defined (for sufficiently nice initial data) in the following
manner:

In a static spacetime, the Klein-Gordon equation (\ref{kg}) can be
written in the form
\begin{equation}
 \partial^2 \Phi / \partial t^2 = - A \Phi \,,
\label{kg2}
\end{equation}
where $t$ denotes the Killing parameter and 
\begin{equation}
 A = - VD^a(VD_a) + m^2 V^2 \,, 
\label{A}
\end{equation}
where $V = (-t^a t_a)^{1/2}$ and $D_a$ is the derivative operator on
$\Sigma$ associated with the induced metric on $\Sigma$. View $A$ as
an operator (with domain $C_0^\infty(\Sigma)$) on the Hilbert space
${\mathcal H} = L^2(\Sigma, \mu)$, where the measure $\mu$ on $\Sigma$
is chosen to be $V^{-1}$ times the natural volume element on $\Sigma$
associated with the induced metric. Then $A$ is a positive, symmetric
operator on $\mathcal H$. Consequently, it admits at least one
positive self-adjoint extension---namely, the Friedrichs extension.
Choose some positive, self-adjoint extension, $A_E$. Given 
$(\phi_0, \dot{\phi}_0) \in C^\infty_0(\Sigma)
\times C^\infty_0(\Sigma)$, for each $t \in {\bf R}$ we define
\begin{equation}
\phi_t = \cos(A_E^{1/2} t) \phi_0 
         + A_E^{-1/2} \sin(A_E^{1/2} t) \dot{\phi}_0 \,. 
\label{dyn}
\end{equation}
Here $\cos(A_E^{1/2} t)$ and $A_E^{-1/2} \sin(A_E^{1/2} t)$ are
bounded operators defined via the spectral theorem. Since, clearly, we
have $(\phi_0, \dot{\phi}_0) \in \mathcal H \times \mathcal H$, it
follows that for all $t$, $\phi_t$ is a well defined element of
$\mathcal H$. It was shown in \cite{w1} that there exists a unique
$\Phi \in C^\infty(M)$ such that for all $t$, $\Phi|_{\Sigma_t} =
\phi_t$ and $t^a \nabla_a \Phi|_{\Sigma_t} = \dot{\phi}_t$, where
$\Sigma_t$ denotes the time translate of $\Sigma$ by $t$. Furthermore,
$\Phi$ satisfies eq.~(\ref{kg}) throughout $M$, and satisfies the
additional properties enumerated in the next section. Therefore,
eq.~(\ref{dyn}) provides a satisfactory prescription for defining the
dynamics associated with eq.~(\ref{kg}), at least for initial data in
$C^\infty_0(\Sigma) \times C^\infty_0(\Sigma)$.

Note that different choices of self-adjoint extensions, $A_E$, give
rise to inequivalent prescriptions for defining dynamical evolution.
(In essence, they correspond to inequivalent choices of ``boundary
conditions at the singularity''.) Thus, if $A$ has more than one
positive self-adjoint extension, there will be more than one choice of
dynamical evolution law as defined by the prescription
eq.~(\ref{dyn}). On the other hand, if $A$ has a unique self-adjoint
extension---as occurs in some cases of interest \cite{hm,ih}---the
general prescription eq.~(\ref{dyn}) yields unique evolution. However,
the analysis of \cite{w1} leaves open the possibility that, in
addition to the freedom of choosing different self-adjoint extensions
of $A$, there could exist other acceptable prescriptions for defining
dynamics that are not even of the form of eq.~(\ref{dyn}). If that
were the case, one could not conclude that dynamical evolution is
uniquely defined even in situations where $A$ has a unique
self-adjoint extension.

The main purpose of this paper is to give a general analysis of the
possibilities for defining dynamics in static, non-globally-hyperbolic
spacetimes, and show that---subject to certain requirements---the
above prescription eq.~(\ref{dyn}) is the only possible one. In the
next section, we shall specify the requirements that we impose upon
the dynamics. In section 3, we shall then prove that these
requirements imply that the dynamics is of the form
eq.~(\ref{dyn}). Applications to defining dynamics in anti-de Sitter
spacetime will be given in a subsequent paper.

\section{Assumptions Concerning Dynamics}

Let $(M, g_{ab})$ be a static (but non-globally-hyperbolic) spacetime,
i.e., $(M, g_{ab})$ possesses an everywhere timelike hypersurface
orthogonal Killing field, $t^a$, whose orbits are complete. We further
assume that there exists a slice $\Sigma$ orthogonal to $t^a$ such
that every orbit of $t^a$ intersects $\Sigma$ once and only once. Let
$\Sigma_t$ denote image of $\Sigma$ under the static isometry by
parameter $t$; it follows that $\{\Sigma_t\}$ provides a foliation of
$M$. We shall label each point $p \in M$ by $(t,x)$, where $x \in \Sigma$
denotes the intersection with $\Sigma$ of the orbit of $t^a$ through
$p$, and $t$ denotes the parameter of $p$ along this orbit starting
from $x$. It follows that $t$ is a global time function on $M$, and hence
$(M, g_{ab})$ is stably causal.

We consider the Klein-Gordon equation (\ref{kg}), which, as noted
above, can be written in the form eq.~(\ref{kg2}), with $A$ being the
differential operator given by eq.~(\ref{A}). We seek a rule for
providing us with a solution to eq.~(\ref{kg}) associated to any
allowed initial data on $\Sigma$. We do not know in advance the widest
possible class of initial data that should be ``allowed'', but we
demand that it at least include all data on $\Sigma$ that is smooth and of
compact support. Thus, our goal is the following: We seek a suitable
prescription such that given any $(\phi_0, \dot{\phi}_0) \in
C^\infty_0(\Sigma) \times C^\infty_0(\Sigma)$, we obtain a unique
$\Phi \in C^\infty(M)$ such that $\Phi$ depends linearly on $(\phi_0,
\dot{\phi}_0)$, satisfies eq.~(\ref{kg}) throughout $M$, and is such
that $\Phi|_\Sigma = \phi_0$ and $t^a \nabla_a \Phi|_\Sigma =
\dot{\phi}_0$. We now state the additional requirements that we
impose upon the prescription so that it is ``suitable''.

First, we require that the solution $\Phi$ must be compatible with
causality in the following sense. 

\medskip
\noindent
{\bf Assumption 1} (causality): Let $K_0$ denote the support of the
initial data on $\Sigma$, i.e., let
\begin{equation}
K_0 = {\rm supp}(\phi_0) \cup {\rm supp}(\dot{\phi}_0) \,. 
\label{K0}
\end{equation}
Then we require that 
\begin{equation}
{\rm supp}(\Phi) \subset J^+(K_0) \cup J^-(K_0) \,,  
\label{caus}
\end{equation}
where $J^{\pm}$ denotes the causal future/past. 
\medskip

Let $\Phi$ be the solution corresponding to the initial data $(\phi_0,
\dot{\phi}_0) \in C^\infty_0(\Sigma) \times C^\infty_0(\Sigma)$. Define
\begin{equation}
\phi_t \equiv \Phi|_{\Sigma_t} \,,  
\label{phit}
\end{equation}
and
\begin{equation}
\dot{\phi}_t \equiv (\partial \Phi/\partial t)|_{\Sigma_t} \,. 
\label{dotphit}
\end{equation}
The following lemma will be useful in many places in our analysis.

\medskip
\noindent
{\bf Lemma 2.1}: Suppose that the dynamics satisfies assumption 1.
Then there exists a $\delta > 0$ such that $\phi_t$ and
$\dot{\phi}_t$ are each in $C^\infty_0(\Sigma_t)$ for all $|t| <
\delta$.

\smallskip
\noindent
{\bf Proof}: Since $\Phi$ is smooth, clearly $\phi_t$ and
$\dot{\phi}_t$ are smooth for all $t$, so we need only show that they
have compact support for sufficiently small $t$.  We show first that
there exists a $\delta > 0$ such that $K_0 \subset {\rm
int}D(\Sigma_t)$ for all $|t| < \delta$, where $D$ denotes the domain
of dependence.  Namely, if not, we could find a sequence $\{x_n\} \in
K_0$ such that $x_n \notin {\rm int}D(\Sigma_{t_n})$ for all $n$,
where the sequence $\{t_n\}$ converges to zero. Since $K_0$ is
compact, there exists an accumulation point $x \in K_0$, and it
follows that $x \notin {\rm int}D(\Sigma_{t_n})$ for sufficiently
large $n$. However, this is impossible, since clearly $x \in {\rm
int}D(\Sigma)$ and the time translation isometries are a continuous
map from ${\bf R} \times M$ into $M$, so we must have $x \in {\rm
int}D(\Sigma_t)$ for sufficiently small $t$.

Now choose $|t| < \delta$; for definiteness we assume that $t \geq
0$. Then, by the above result, we have $J^+(K_0) \cap J^-(\Sigma_t)
\subset {\rm int}D(\Sigma_t)$, so we may restrict attention to the
globally hyperbolic sub-spacetime ${\rm int}D(\Sigma_t)$. But, since
$K_0$ is compact, it follows from a slight generalisation of a
standard theorem (see theorem 8.3.12 of \cite{w}) that $J^+(K_0) \cap
\Sigma_t$ is compact. It then follows immediately from assumption 1
above that the supports of $\phi_t$ and $\dot{\phi}_t$ are compact. $\Box$

\medskip
Our second set of requirements ensure that our prescription is
compatible with the spacetime symmetries. Let $T_t: C^\infty(M)
\rightarrow C^\infty(M)$ denote the natural action of the time
translation isometries on smooth functions on $M$, i.e., for any
smooth $F:M \rightarrow M$ we define
\begin{equation}
(T_t F)(s,x) = F(s-t,x) \,. 
\label{Tt}
\end{equation}
Similarly, let $P: C^\infty(M) \rightarrow C^\infty(M)$ denote the
natural action of the time reflection isometry on smooth functions on $M$,
i.e., for any smooth $F:M \rightarrow M$ we define
\begin{equation}
(P F)(t,x) = F(-t,x) \,. 
\label{P}       
\end{equation}
We require our prescription to satisfy the following properties: 

\medskip 
\noindent 
{\bf Assumption 2(i)} (time translation invariance):
Let $\Phi$ be the solution associated with the data $(\phi_0,
\dot{\phi}_0)$ on $\Sigma$. Suppose that $K_t \equiv {\rm supp}(\phi_t)
\cup {\rm supp}(\dot{\phi}_t)$ is compact---which, as shown above, it
must be for all $|t| < \delta$ for some $\delta > 0$. We require that
if the ``time translated'' data $(\phi_t, \dot{\phi}_t)$ is specified
on $\Sigma$, then the corresponding solution must be the time
translate, $T_{-t} \Phi$, of the original solution
$\Phi$. Furthermore, if the data $(\dot{\phi}_0, -A \phi_0)$ is
specified on $\Sigma$---which formally corresponds to initial data for
$\partial \Phi/\partial t$---then we demand that the corresponding
solution is $\partial \Phi/\partial t$. 
\medskip

\noindent
{\bf Assumption 2(ii)} (time reflection invariance): If the data $(\phi_0,
-\dot{\phi}_0)$ is specified on $\Sigma$---which formally corresponds
to initial data for the time reverse of $\Phi$---then we require that
the corresponding solution is $P \Phi$.
\medskip

Our final set of requirements concern the existence of a suitable
conserved ``energy'' for solutions. Consider, first, the case of a
globally hyperbolic, static spacetime. In this case, if $(\phi_0,
\dot{\phi}_0)$ is of compact support, then $(\phi_t, \dot{\phi}_t)$
will be of compact support for all $t$. For the unique solution,
$\Phi$, corresponding to this initial data, define
\begin{eqnarray}
E(\Phi, \Phi) &=& \int_\Sigma \dot{\phi}_0^2 V^{-1}d\Sigma +
\int_\Sigma \phi_0 A \phi_0 V^{-1}d\Sigma \nonumber \\ 
&=& (\dot{\phi}_0, \dot{\phi}_0)_{L^2} + (\phi_0, A\phi_0)_{L^2} \,,  
\label{E}
\end{eqnarray}
where the $L^2$ inner product is defined using the volume element
specified below eq.~(\ref{A}), and the integrals clearly converge
since all functions appearing in the integrals are smooth and of
compact support. It is easily seen that $E$ is positive
definite, and it follows immediately from the Klein-Gordon equation
(\ref{kg2}) that $E$ is conserved in the sense that
\begin{equation}
E(T_t \Phi, T_t \Phi) = E(\Phi, \Phi) \,. 
\label{Econs}
\end{equation}
We may therefore view $E$ as a conserved inner product on the
vector space, ${\cal W}$ of solutions with initial data in
$C^\infty_0(\Sigma) \times C^\infty_0(\Sigma)$. More generally, if
$\Phi$ is a solution with initial data in $C^\infty_0(\Sigma) \times
C^\infty_0(\Sigma)$ and $\Xi$ is a solution with initial data that is
merely in $C^\infty(\Sigma) \times C^\infty(\Sigma)$, then
\begin{equation}
E(\Xi, \Phi) = (\dot{\xi}_0, \dot{\phi}_0)_{L^2} + (\xi_0, A
\phi_0)_{L^2}
\label{E2}
\end{equation}
also is well defined and is conserved. 

We shall require our prescription for defining dynamics in the
non-globally-hyperbolic case to admit a conserved energy which reduces
to eq.~(\ref{E2}) in appropriate cases. In the non-globally-hyperbolic
case, a solution with data $(\phi_0, \dot{\phi}_0)$ in
$C^\infty_0(\Sigma) \times C^\infty_0(\Sigma)$ will not, in general,
be such that $\phi_t$ and $\dot{\phi}_t$ will be of compact support
for all $t$ (although, as shown above, $\phi_t$ and $\dot{\phi}_t$
will be of compact support for all $|t| < \delta$ for some $\delta >
0$). Consequently, if we define $\cal W$ to be the vector space of
solutions (as given by our prescription) with initial data $(\phi_0,
\dot{\phi}_0)$ in $C^\infty_0(\Sigma) \times C^\infty_0(\Sigma)$, then
$T_t$ will not map $\cal W$ into itself.  It is therefore useful to
enlarge the solution space $\cal W$ as follows. 

Let $\cal V$ denote the vector space of solutions to the Klein-Gordon
equation that can be expressed as finite linear combinations of
solutions of the form $T_t \Phi$ for $\Phi \in {\cal W}$ and $t \in
{\bf R}$, i.e.,
\begin{equation}
{\cal V} \equiv \{\Psi| \Psi = T_{t_1} \Phi_1 + ... + T_{t_n} \Phi_n,
\Phi_i \in {\cal W}\} \,. 
\label{V}  
\end{equation}   
(Note that in the globally hyperbolic case, we have ${\cal V} = {\cal
W}$, but in the non-globally-hyperbolic case, in general, ${\cal W}$
will be a proper subset of ${\cal V}$.) Define the initial data
spaces, $X$ and $Y$, for $\cal V$ by
\begin{equation}
X \equiv \{\psi_0 = \Psi|_\Sigma | \Psi \in {\cal V}\} \,,  
\label{X}
\end{equation}
\begin{equation}
Y \equiv \{\dot{\psi}_0 = (\partial \Psi/\partial t)|_\Sigma | 
  \Psi\in {\cal V}\} \,. 
\label{Y}
\end{equation}
Clearly, we have $C^\infty_0(\Sigma) \subset X$ and
$C^\infty_0(\Sigma) \subset Y$. Several additional properties of the
spaces $X$ and $Y$ follow immediately from our previous
assumptions. First, since the last condition of assumption 2(i)
implies that if $\Psi \in {\cal V}$ then $\partial \Psi/\partial t \in
{\cal V}$, it follows immediately that $Y \subset X$. Furthermore,
assumption 2(ii) implies that if $\Psi \in {\cal V}$ then $P \Psi \in
{\cal V}$, and consequently, $(\Psi \pm P \Psi) \in {\cal V}$. It
follows that if $(\psi_0, \dot{\psi}_0)$ is initial data for a
solution $\Psi$ in $\cal V$, then $(\psi_0, 0)$ and $(0,
\dot{\psi}_0)$ are also initial data for a solution $\cal V$.
Consequently, the vector space of initial data for solutions in $\cal
V$ is isomorphic to $X \oplus Y$.

We require that our prescription for dynamics be such that there exist
a symmetric, positive definite bilinear map (i.e., an inner product)
$E: {\cal V} \times {\cal V} \rightarrow {\bf R}$---which we may
equivalently view as a bilinear map from $(X \oplus Y) \times (X
\oplus Y)$ into ${\bf R}$---such that the following properties hold:

\medskip
\noindent
{\bf Assumption 3(i)} (time translation invariance of $E$): $E$ is
time translation invariant in the sense that for all $\Psi_1, \Psi_2
\in {\cal V}$ and all $t$ we have
\begin{equation}
E(T_t \Psi_1, T_t \Psi_2) = E(\Psi_1, \Psi_2) \,.
\label{Econs3}
\end{equation}

\medskip 
\noindent 
{\bf Assumption 3(ii)} (time reflection invariance of $E$): $E$ is
time reflection invariant in the sense that for all $\Psi_1, \Psi_2
\in {\cal V}$ we have
\begin{equation}
E(P \Psi_1, P \Psi_2) = E(\Psi_1, \Psi_2)  \,.  
\label{Econs4}
\end{equation}

\medskip
\noindent
{\bf Assumption 3(iii)} (agreement with formula in the globally
hyperbolic case): If $\Phi \in {\cal W}$ and $\Psi \in {\cal V}$ (but
$\Psi$ need not be in $\cal W$) then $E$ is given the same formula as
in the globally hyperbolic case (see eq.~(\ref{E2}) above), i.e.,
\begin{equation}
E(\Psi, \Phi) = (\dot{\psi}_0, \dot{\phi}_0)_{L^2} + (\psi_0, A
                 \phi_0)_{L^2} \,.
\label{E3}
\end{equation}

\medskip
\noindent
{\bf Assumption 3(iv)} (compatibility of convergence with respect to 
$E$ with more elementary notions of convergence of initial data): 
Suppose that $\{\Psi_n\}$ is a sequence in $\cal V$ that is a Cauchy
with respect to the norm defined by $E$. Suppose, further, that there
exists a $\Psi \in {\cal V}$ such that $(\psi_n)_0$ and all of its
spatial derivatives converge uniformly on compact subsets of $\Sigma$
to $\psi_0$ and its corresponding spatial derivatives and, similarly,
that $(\dot{\psi}_n)_0$ and all of its spatial derivatives converge
uniformly on compact subsets of $\Sigma$ to $\dot{\psi}_0$ and its
corresponding spatial derivatives. Then we require that $\{\Psi_n\}$
converge to $\Psi$ in the norm defined by $E$, i.e., we require that
\begin{equation}
\lim_{n \rightarrow \infty} E(\Psi_n - \Psi, \Psi_n - \Psi) = 0 \,.
\end{equation}

\section{General Analysis of Dynamics}

In this section, we shall prove that any prescription for defining
dynamics that satisfies the assumptions of the previous section must,
in fact, be equivalent to the prescription of \cite{w1}, obtained by
choosing a self-adjoint extension of the operator $A$ on $L^2(\Sigma,
\mu)$.

It will be convenient to view the energy, $E$, introduced in the
previous section, as a bilinear map $E: (X \oplus Y) \times (X \oplus
Y) \rightarrow {\bf R}$ on initial data space. We may break $E$ up
into a sum of four maps of the form $Q:X \times X \rightarrow {\bf
R}$, $S:Y \times Y \rightarrow {\bf R}$, $R:X \times Y \rightarrow
{\bf R}$, and $T:Y \times X \rightarrow {\bf R}$, i.e., we may write
\begin{equation}
E([\psi_0, \dot{\psi}_0], [\chi_0, \dot{\chi}_0]) = Q(\psi_0, \chi_0)
+ S(\dot{\psi}_0, \dot{\chi}_0) + R(\dot{\psi}_0, \chi_0) + T(\psi_0,
\dot{\chi}_0) \,. 
\label{E4}
\end{equation}
By assumption 3(ii), $E$ is invariant under the time reflection map
$P$. Since the action of $P$ on initial data is to take $(\psi_0,
\dot{\psi}_0)$ into $(\psi_0, - \dot{\psi}_0)$, it follows immediately
that the time reflection invariance of $E$ implies that $R=T=0$. The
positive definiteness of $E$ on $\cal V$ implies that $Q$ defines an
inner product on $X$ and that $S$ defines an inner product on $Y$. In
fact, we now show that the inner product, $S$, on $Y$
is just the $L^2$ inner product considered in the analysis 
of~\cite{w1}. 

\medskip
\noindent
{\bf Lemma 3.1}: Let $Y$ and $S: Y \times Y \rightarrow {\bf R}$ be
defined as above (see eqs.~(\ref{Y}) and (\ref{E4})). Then, under the
assumptions stated in the previous section, we have $Y \subset
L^2(\Sigma, \mu)$, where the measure, $\mu$ is that arising from
$V^{-1}$ times the natural volume element on $\Sigma$. Furthermore,
for all $\xi, \eta \in Y$, we have
\begin{equation}
S(\xi, \eta) = (\xi, \eta)_{L^2} \,.  
\label{S}
\end{equation}

\smallskip
\noindent
{\bf Proof}: Let $\xi \in Y$ and consider the solution $\Psi \in \cal
V$ with initial data $(\psi_0 = 0, \dot{\psi}_0 = \xi)$. Let $\{O_n\}$
be a nested family of open subsets of $\Sigma$ with compact closure
such that $\cup_n O_n = \Sigma$. (Such a family can be constructed by
putting a complete Riemannian metric on $\Sigma$, choosing a point $p
\in \Sigma$, and taking $O_n$ to be the open ball of radius $n$ in
this metric about point $p$.) Let $\{f_n\}$ be a sequence of functions
in $C^\infty_0(\Sigma)$ with $0 \leq f_n \leq 1$ everywhere, and
$f_n(x) = 1$ if $x \in O_{n-1}$ but $f_n(x) = 0$ if $x \notin
O_{n+1}$. Let $\xi_n = f_n \xi$. Let $\Psi_n \in {\cal W}$ be the
solution associated with the initial data $(0, \xi_n)$. Then we have 
\begin{eqnarray}
S(\xi, \xi) &=& E(\Psi, \Psi) \nonumber \\ 
&=& E(\Psi - \Psi_n, \Psi - \Psi_n) +
E(\Psi - \Psi_n, \Psi_n) \nonumber \\
&& + E(\Psi_n, \Psi - \Psi_n) + E(\Psi_n,
\Psi_n) \nonumber \\ 
&=& E(\Psi - \Psi_n, \Psi - \Psi_n) + 2\int_\Sigma f_n(1-f_n)
\xi^2 V^{-1}d\Sigma \nonumber \\
&& + \int_\Sigma (f_n\xi)^2 V^{-1}d\Sigma \nonumber \\
&\geq& \int_\Sigma (f_n\xi)^2 V^{-1}d\Sigma \,, 
\label{S2}
\end{eqnarray}
where we have used assumption 3(iii) to obtain an explicit form for
$E$ in the case where one of the arguments, $\Psi_n$, lies in $\cal
W$. However, the last inequality in eq.~(\ref{S2}) implies that $\xi
\in L^2(\Sigma, \mu)$, since, if not, given any $C>0$ we could find an
$n$ such that $\|f_n \xi \|^2_{L^2} > C$. This shows that $Y \subset
L^2$. But, since $\xi \in L^2$, it follows that $\{\xi_n = f_n \xi\}$
is a Cauchy sequence in $L^2$, i.e., $\{\Psi_n\}$ is a Cauchy sequence
in the energy norm. Since $\{\xi_n\}$ and all spatial derivatives of
$\{\xi_n\}$ clearly converge uniformly on all compact subsets of
$\Sigma$ to $\xi$ and its corresponding spatial derivatives, it
follows from assumption 3(iv) that $\{\Psi_n\}$ converges to $\Psi$ in
the energy norm. But this means that for all $\xi \in Y$ we have
\begin{equation} 
S(\xi, \xi) = \lim_{n \rightarrow \infty} \int_\Sigma (f_n\xi)^2
V^{-1}d\Sigma = \|\xi \|^2_{L^2} \,, 
\label{S3}
\end{equation}
which implies that the inner product on $Y$ defined by $S$ is just the
$L^2$ inner product, as we desired to show. $\Box$

\medskip

Now complete $\cal V$ in the inner product $E$ to obtain the real Hilbert
space ${\mathcal H}_E$. Similarly, we complete $X$ and $Y$ in the
inner products $Q$ and $S$, respectively, to obtain real Hilbert spaces
${\mathcal H}_X$ and ${\mathcal H}_Y = L^2(\Sigma, \mu)$,
respectively. Clearly, we have
\begin{equation}
{\mathcal H}_E \cong {\mathcal H}_X \oplus L^2(\Sigma, \mu) \,. 
\label{EXY}
\end{equation}

Now, by assumption 3(i), for each $t$, $T_t: {\cal V} \rightarrow
{\cal V}$ is a norm preserving---and, hence, bounded---linear map in
the norm $E$, so it can be uniquely extended to a bounded linear map
$T_t: {\mathcal H}_E \rightarrow {\mathcal H}_E$. By continuity, the
extended $T_t$ also is norm preserving. Furthermore, the range of
$T_t$ clearly includes $\cal V$, so the range of $T_t$ is dense in
${\mathcal H}_E$. Consequently, $T_t: {\mathcal H}_E \rightarrow
{\mathcal H}_E$ is unitary for each $t$. Furthermore, on $\cal V$ we
have $T_t \circ T_s = T_{t+s}$ and, by continuity, this relation must
hold on ${\mathcal H}_E$. Thus, $T_t$ is a one-parameter unitary group
on ${\mathcal H}_E$. Furthermore, we have the following proposition:

\medskip
\noindent
{\bf Proposition 3.1}: For all $\Psi \in {\cal V}$, $T_t \Psi$ is
strongly differentiable in ${\mathcal H}_E$, and its derivative is
$\partial \Psi/\partial t$.

\smallskip  
\noindent
{\bf Proof}: Our task is to show that for all $\Psi \in {\cal V}$,
\begin{equation}
\lim_{t \rightarrow 0} \left\|\frac{T_t \Psi - \Psi}{t} - \frac{\partial
\Psi}{\partial t} \right\|_E = 0 \,.
\label{diff}
\end{equation}
Since $\cal V$ consists of finite sums of time translates of elements
of $\cal W$ (see eq.~(\ref{V}) above), it suffices to prove
eq.~(\ref{diff}) for solutions of the form $T_s \Phi$ for $\Phi \in
{\cal W}$. However, since $T_t$ is a unitary group, it follows
immediately that it suffices to prove eq.~(\ref{diff}) for $\Phi \in
{\cal W}$. In that case, we also have $\partial \Phi/\partial t \in
{\cal W}$ and, by Lemma 2.1, there exists a $\delta > 0$ such that
$T_t \Phi \in {\cal W}$ for all $|t| < \delta$. Consequently, by
assumption 3(iii), the energy norm appearing in eq.~(\ref{diff}) is
given explicitly by eq.~(\ref{E3}). The result then follows
immediately from standard results for solutions with smooth data of
compact support to the partial differential eq.~(\ref{kg}) on the
globally hyperbolic spacetime ${\rm int} D(\Sigma)$. $\Box$

\medskip
\noindent
{\bf Corollary}: $T_t$ defines a strongly continuous one-parameter
group on ${\mathcal H}_E$.

\smallskip
\noindent
{\bf Proof}: We have already established that $T_t$ is a one-parameter
group on ${\mathcal H}_E$, so we need only show that it is strongly
continuous. However, we have just shown that $T_t \Psi$ is strongly
differentiable---and, hence, strongly continuous---for all $\Psi \in
{\cal V}$. Since ${\cal V}$ is dense in ${\mathcal H}_E$, given $\Xi \in
{\mathcal H}_E$ and given $\epsilon > 0$ we can find a $\Psi \in {\cal
V}$ such that $\|\Xi - \Psi \|_E < \epsilon/3$. Writing
\begin{equation}
T_t \Xi - \Xi = T_t (\Xi - \Psi) - (\Xi - \Psi) + (T_t \Psi - \Psi)
\,, 
\label{xi}
\end{equation}
we see immediately that 
\begin{equation}
\| T_t \Xi - \Xi \|_E \leq 2\epsilon/3 + \|T_t \Psi - \Psi \|_E 
\label{xi2}
\end{equation}
so strong continuity of $T_t$ on $\cal V$ implies strong continuity of
$T_t$ on ${\mathcal H}_E$. $\Box$

\medskip
Since $T_t$ is a strongly continuous one-parameter unitary group on
${\mathcal H}_E$, by Stone's theorem (adapted here to the real Hilbert
space case), there exists a skew adjoint map ${\cal B}: {\mathcal H}_E
\rightarrow {\mathcal H}_E$ such that $T_t = \exp(-t{\cal
B})$. Furthermore, since $T_t \Psi$ is strongly differentiable for all
$\Psi \in {\cal V}$ we have ${\cal V} \subset {\rm Dom}({\cal
B})$. Since we have shown in Proposition 3.1 above that for all $\Psi
\in {\cal V}$ the derivative of $T_t \Psi$ is just $\partial
\Psi/\partial t$, we have for all $\Psi \in {\cal V}$
\begin{equation}
\frac{\partial \Psi}{\partial t} = -{\cal B} \Psi \,. 
\label{B1}
\end{equation}
Using the isomorphism (\ref{EXY}), we can view $\cal B$ as a
skew-adjoint operator on ${\mathcal H}_X \oplus L^2(\Sigma, \mu)$ and
re-write eq.~(\ref{B1}) as the pair of equations
\begin{equation}
\dot{\psi}_0 = - {\cal B}_{11} \psi_0 - {\cal B}_{12} \dot{\psi}_0 \,, 
\label{B2}
\end{equation}
\begin{equation}
\ddot{\psi}_0 = - {\cal B}_{21} \psi_0 - {\cal B}_{22} \dot{\psi}_0
\,, 
\label{B3}
\end{equation}
where 
\begin{equation}
\ddot{\psi}_0 \equiv (\partial^2 \Psi/\partial^2 t)|_{\Sigma} \,, 
\label{ddot}
\end{equation}
and ${\cal B}_{11}: {\mathcal H}_X \rightarrow {\mathcal H}_X$, 
${\cal B}_{12}: L^2(\Sigma, \mu) \rightarrow {\mathcal H}_X$, etc. 

Since the domain of $\cal B$ contains $\cal V$, eqs.~(\ref{B2}) and
(\ref{B3}) are guaranteed to hold for all $\psi_0 \in X$ and all
$\dot{\psi}_0 \in Y$. It follows immediately that the restriction of
${\cal B}_{11}$ to $X$ vanishes and that the restriction of ${\cal
B}_{12}$ to $Y \subset L^2(\Sigma, \mu)$ must equal $-I$, where $I$
denotes the identity map on functions. (Note, however, that ${\cal
B}_{12}$ is a map between different Hilbert spaces.) Similarly, the
invariance of the dynamics under time reflection (assumption 2(ii)
above) implies that the restriction of ${\cal B}_{22}$ to $Y$ must
vanish. Thus, writing $B = {\cal B}_{21}$, we have learned that there
exists an operator $B: {\mathcal H}_X \rightarrow L^2(\Sigma, \mu)$
with $X \subset {\rm Dom}(B)$ such that for all $\psi_0 \in X$ we have
\begin{equation}
\ddot{\psi}_0 = - B \psi_0 \,. 
\label{ddot2}
\end{equation}
Comparing with eq.~(\ref{kg2}), we see that if $\psi_0 \in C^\infty_0
(\Sigma)$, then $B = A$, where $A$ was defined by eq.~(\ref{A}). Thus,
when viewed as an operator from ${\mathcal H}_X$ into $L^2(\Sigma,
\mu)$, $B$ is an extension of $A$ from the domain $C^\infty_0
(\Sigma)$ to a domain that includes all of $X$.

Finally, we note that the skew-adjointness of ${\cal B}$ on ${\mathcal
H}_E$ directly implies that for all $\psi_0, \chi_0 \in X$ and all
$\dot{\psi}_0, \dot{\chi}_0 \in Y$, we have
\begin{equation}
- (\chi_0, \dot{\psi}_0)_{{\mathcal H}_X} + (\dot{\chi}_0, B
\psi_0)_{L^2} = (\psi_0, \dot{\chi}_0)_{{\mathcal H}_X} -
(\dot{\psi}_0, B \chi_0)_{L^2} \,. 
\label{sa}
\end{equation}
But this equation can hold if and only if for all $\psi_0 \in
X$ and all $\dot{\psi}_0 \in Y$ we have
\begin{equation}
(\dot{\psi}_0, B \psi_0)_{L^2} = (\psi_0, \dot{\psi}_0)_{{\mathcal H}_X}
\label{sa2} \,. 
\end{equation}

To make further progress, we need the following lemma:

\medskip
\noindent
{\bf Lemma 3.2}: Let $\Phi \in {\cal W}$ and, as before, write $\phi_t
= \Phi|_{\Sigma_t}$ and $\dot{\phi}_t = (\partial \Phi/\partial
t)|_{\Sigma_t}$. Then $\phi_t \in L^2(\Sigma, \mu)$ (and, hence, $X
\subset L^2(\Sigma, \mu)$). Furthermore, $\phi_t$ is strongly
differentiable with respect to $t$ as a vector in $L^2(\Sigma, \mu)$
and $d\phi_t/dt = \dot{\phi}_t$. More generally, $d^n \phi_t/dt^n$
exists and equals $(\partial^n \Phi/\partial t^n)|_{\Sigma_t}$.

\smallskip
\noindent
{\bf Remark}: We previously showed that $Y \subset L^2(\Sigma, \mu)$
and also that $Y \subset X$. In addition, we previously showed that
$T_t \Psi$ is strongly differentiable with respect to $t$ as a vector
in ${\mathcal H}_E$. None of the results claimed in Lemma 3.2 follow
immediately from these previous results.

\smallskip
\noindent
{\bf Proof}: We label points by $(t,x)$ as explained at the beginning
of section 2. Then at each fixed $x$ we have
\begin{equation}
\phi_t - \phi_0 = \int_0^t \dot{\phi}_{t'} dt' = \int_0^t [\dot{\phi}_0
+ (\dot{\phi}_{t'} - \dot{\phi}_0)] dt' \,,  
\label{phit1}
\end{equation}
and hence
\begin{equation}
\phi_t - \phi_0 - t \dot{\phi}_0 =\int_0^t (\dot{\phi}_{t'} -
\dot{\phi}_0) dt' \,. 
\label{phit2}
\end{equation}
It follows that
\begin{eqnarray}
\left|\phi_t - \phi_0 - t \dot{\phi}_0 \right|^2 
   &=& \left| \int_0^t (\dot{\phi}_{t'} 
             - \dot{\phi}_0) dt' 
       \right|^2 \nonumber \\ 
&\leq& \int_0^t |1|^2 dt' \int_0^t \left|\dot{\phi}_{t'} -
\dot{\phi}_0 \right|^2 dt' \nonumber \\
&=& t \int_0^t \left|\dot{\phi}_{t'} - \dot{\phi}_0 \right|^2 dt' \,,
\label{phit3}
\end{eqnarray}
where the Schwartz inequality was used in the second line.

Now let $\{f_n\}$ be as in the proof of Lemma 3.1 above. We multiply
eq.~(\ref{phit3}) by $f^2_n$ and integrate over $\Sigma$ with respect to
the volume element $V^{-1} d\Sigma$. We obtain
\begin{eqnarray}
\left\|f_n(\phi_t - \phi_0 - t \dot{\phi}_0) \right\|^2_{L^2} 
 &\leq& t \int_\Sigma V^{-1} d\Sigma \int_0^t dt' 
          \left|f_n(\dot{\phi}_{t'} - \dot{\phi}_0) \right|^2 \nonumber \\ 
&=& t \int_0^t dt' \int_\Sigma V^{-1} d\Sigma 
          \left| f_n(\dot{\phi}_{t'} - \dot{\phi}_0) \right|^2 \nonumber \\
&\leq& t \int_0^t dt' \int_\Sigma V^{-1} d\Sigma 
          \left|\dot{\phi}_{t'} - \dot{\phi}_0 \right|^2 \nonumber \\
&\leq& t \int_0^t dt' E(T_{-t'}\Phi - \Phi, T_{-t'} \Phi - \Phi) \nonumber \\
&\leq& 4t \int_0^t dt E(\Phi, \Phi) \nonumber \\ 
&=& 4t^2 E(\Phi, \Phi) \,. 
\label{phit4}
\end{eqnarray}
By the same argument as used in Lemma 3.1, this inequality shows that
$(\phi_t - \phi_0 - t \dot{\phi}_0) \in L^2(\Sigma, \mu)$. However,
since $\phi_0$ and $\dot{\phi}_0$ are in $C^\infty_0(\Sigma)$, this
proves that $\phi_t \in L^2(\Sigma, \mu)$, as we desired to show.

To prove the differentiability of $\phi_t$, we need to
generalise and sharpen our above estimates. First, we repeat the same
steps as led to eq.~(\ref{phit4}) but now use arbitrary times $t'$ and
$t$ rather than $t$ and $0$ to find that for all $t$ and all $t'$ we
have
\begin{equation}
\left\|\phi_{t'} - \phi_{t} - (t'-t) \dot{\phi}_t \right\|^2_{L^2} 
 \leq 4(t'-t)^2 E(\Phi, \Phi) \,. 
\label{phit5}
\end{equation}
Applying this result to the solution $\partial \Phi/\partial t$, we obtain
\begin{equation}
\left\|\dot{\phi}_{t'} - \dot{\phi}_{t} 
       - (t'-t) \ddot{\phi}_t \right\|^2_{L^2} 
\leq 4(t'-t)^2 E(\partial \Phi/\partial t, \partial \Phi/\partial t) \,,
\label{phit6}
\end{equation}
which implies that at any fixed $t$, there exists a $C > 0$ and a
$\delta > 0$ such that for all $|t'-t| < \delta$, we have
\begin{equation}
\left\|\dot{\phi}_{t'} - \dot{\phi}_{t} \right\|_{L^2} \leq C |t'-t| \,.
\label{phit7}
\end{equation}
On the other hand, now that we know that $\phi_t \in L^2(\Sigma, \mu)$ 
for all $t$, the same steps as led to the third line of
eq.~(\ref{phit4}) now yield 
\begin{equation}
\left\|\phi_{t'} - \phi_{t} - (t'-t) \dot{\phi}_t \right\|^2_{L^2} 
\leq (t'-t)
\int_t^{t'} dt'' \left\|\dot{\phi}_{t''} 
    - \dot{\phi}_{t} \right\|^2_{L^2}  \,. 
\label{phit8} 
\end{equation} 
Substituting the estimate (\ref{phit7}), we find that 
for all $|t'-t|< \delta$, we have 
\begin{equation}
\left\|\phi_{t'} - \phi_{t} - (t'-t) \dot{\phi}_t \right\|^2_{L^2} 
  \leq C' (t'-t)^4 \,. 
\label{phit9}
\end{equation}
Dividing this equation by $(t'-t)^2$ and taking the limit $t'
\rightarrow t$, we immediately see that the strong derivative of
$\phi_t$ in $L^2(\Sigma, \mu)$ exists and is equal to $\dot{\phi}_t$.
The results for higher time derivatives of $\phi_t$ follow immediately
by applying the same arguments to $\partial^n \Phi/\partial t^n$. $\Box$

\medskip

We now prove our main theorem:

\medskip
\noindent
{\bf Theorem 3.1}: Consider any prescription for assigning a solution,
$\Phi$, to any initial data $(\phi_0, \dot{\phi}_0) \in
C^\infty_0(\Sigma) \times C^\infty_0(\Sigma)$ such that all of the
assumptions stated in section 2 are satisfied. Let $A: L^2(\Sigma,
\mu) \rightarrow L^2(\Sigma, \mu)$ be the operator (\ref{A}) defined
on the domain $C^\infty_0(\Sigma)$. Then there exists a positive,
self-adjoint extension, $A_E: L^2(\Sigma, \mu) \rightarrow L^2(\Sigma,
\mu)$, of $A$, such that the dynamics defined by eq.~(\ref{dyn}) agrees
with the dynamics given by the prescription.

\smallskip
\noindent
{\bf Proof}: We have already learned that for all $\Psi \in {\cal V}$,
there exists an operator $B: {\mathcal H}_X \rightarrow L^2(\Sigma, 
\mu)$ such that eq.~(\ref{ddot2}) holds. Furthermore, we know that $B$
is an extension of $A$ to a domain that contains $X$ and is such that
eq.~(\ref{sa2}) holds for all $\psi \in X$ and $\dot{\psi}_0 \in
Y$. We also know that $Y \subset X \subset L^2(\Sigma, \mu)$. 

Now let $C$ denote the restriction of $B$ to the domain $Y$ and view
$C$ as a map $C: L^2(\Sigma, \mu) \rightarrow L^2(\Sigma, \mu)$ rather
than a map of ${\mathcal H}_X$ into $L^2(\Sigma, \mu)$. Since
$C^\infty_0(\Sigma) \subset Y$, we see that $C$ is densely defined on
$L^2(\Sigma, \mu)$ and eq.~(\ref{sa2}) shows that, when viewed as an
operator on $L^2(\Sigma, \mu)$, $C$ is positive and symmetric on the
domain $Y$. Therefore, there exists a positive, self-adjoint extension
of $C$, which we shall denote as $D: L^2(\Sigma, \mu) \rightarrow
L^2(\Sigma, \mu)$.

Consider, first, a solution $\Phi \in {\cal W}$ which is of the form
$\Phi = \partial \Xi/\partial t$ for some $\Xi \in {\cal W}$. Then for
all $t$ we have $\phi_t = \dot{\xi}_t$ so $\phi_t \in Y$ (and, of
course, by definition of $Y$, we have $\dot{\phi}_t \in Y$ for all
$t$). For each $t$, define
\begin{equation}
\sigma_t = \cos(D^{1/2} t) \phi_t - D^{-1/2} \sin(D^{1/2} t)
\dot{\phi}_t \,. 
\label{dyn2}
\end{equation}
Note that this formula corresponds to using the given prescription to
evolve forward by $t$ to obtain $(\phi_t, \dot{\phi}_t)$, and then
using the prescription (\ref{dyn}) with $A_E = D$ to evolve {\it
backwards} by $t$. Since by Lemma 3.2 $\phi_t$ and $\dot{\phi}_t$ are
strongly differentiable in $t$ and since $\phi_t \in Y \subset 
{\rm Dom}(D) \subset {\rm Dom}(D^{1/2})$, it follows that $\sigma_t$
is strongly differentiable in $t$ and its derivative as a vector in 
$L^2(\Sigma, \mu)$ is given by
\begin{eqnarray}
\frac{d \sigma_t}{dt} = &-& D^{1/2} \sin(D^{1/2} t) \phi_t +
\cos(D^{1/2} t) \frac{d \phi_t}{dt} \nonumber \\
&-& \cos(D^{1/2} t) \dot{\phi}_t -
D^{-1/2} \sin(D^{1/2} t) \frac{d \dot{\phi}_t}{dt} \,. 
\label{dyn3}
\end{eqnarray}
However, by Lemma 3.2 we have $d\phi_t/dt = \dot{\phi}_t$. Furthermore,
we have
\begin{equation}
\frac{d \dot{\phi}_t}{dt} 
 = \ddot{\phi}_t = -B \phi_t = - D \phi_t \,, 
\label{ddphi}
\end{equation}
where the first equality follows from Lemma 3.2, the second from
eq.~(\ref{ddot2}), and the third from the fact that $D$ agrees with
$B$ for all vectors in $Y$. We therefore obtain $d \sigma_t/dt = 0$
and thus $\sigma_t = \sigma_0 = \phi_0$, i.e., we have found that 
\begin{equation}
\phi_0 = \cos(D^{1/2} t) \phi_t - D^{-1/2} \sin(D^{1/2} t)
\dot{\phi}_t \,. 
\label{dyn4}
\end{equation}
In a similar manner, it follows that 
\begin{equation}
\dot{\phi}_0 = \cos(D^{1/2} t) \dot{\phi}_t - D^{-1/2} \sin(D^{1/2} t)
(-D \phi_t) \,. 
\label{dyn5}
\end{equation}
These equations can be inverted to yield
\begin{equation}
\phi_t = \cos(D^{1/2} t) \phi_0 + D^{-1/2} \sin(D^{1/2} t)
\dot{\phi}_0 \,. 
\label{dyn6}
\end{equation}
This proves that all solutions of the form $\Phi = \partial
\Xi/\partial t$ are as claimed in the theorem, with $A_E = D$.

Now consider a solution $\Phi \in {\cal W}$ arising from arbitrary
initial data $(\phi_0, \dot{\phi}_0) \in C^\infty_0(\Sigma) \times
C^\infty_0(\Sigma)$. Let $\tilde{\Phi}$ denote the solution defined by
eq.~(\ref{dyn}) with $A_E = D$. It was proven in \cite{w1} that
$\tilde{\Phi}$ is a smooth solution to the Klein-Gordon equation
(\ref{kg}) with initial data $(\phi_0, \dot{\phi}_0)$. It also follows
from the analysis of \cite{w1} that $d \tilde{\phi}_t/dt$ exists (in
the strong Hilbert space sense in $L^2(\Sigma, \mu)$) and equals
$(\partial \tilde{\Phi}/\partial t)|_{\Sigma_t}$, where
$\tilde{\phi}_t \equiv \tilde{\Phi}|_{\Sigma_t}$. However,
differentiation of eq.~(\ref{dyn}) yields
\begin{equation}
\frac {d \tilde{\phi}_t}{dt} = \cos(D^{1/2} t) \dot{\phi}_0 + D^{-1/2}
\sin(D^{1/2} t) (-D \phi_0) \,. 
\label{dyn7}
\end{equation}
Eq.~(\ref{dyn7}) shows that $\partial \tilde{\Phi}/\partial t$ is the
solution given by eq.~(\ref{dyn}) with $A_E = D$ for the initial data
$(\dot{\phi}_0, - A \phi_0)$. However, by assumption 2(i), the
solution with initial data $(\dot{\phi}_0, - A \phi_0)$ as given by
our prescription is $\partial \Phi/\partial t$. Furthermore, we have
just shown above that for solutions of the form $\partial
\Phi/\partial t$, our prescription must agree with the prescription
given by eq.~(\ref{dyn}) with $A_E = D$. It therefore follows that
$\partial \tilde{\Phi}/\partial t = \partial \Phi/\partial
t$. However, since $\tilde{\Phi}|_\Sigma = \Phi|_\Sigma$, it follows
immediately that $\tilde{\Phi} = \Phi$, i.e., our prescription for
obtaining a solution associated with arbitrary data $(\phi_0,
\dot{\phi}_0) \in C^\infty_0(\Sigma) \times C^\infty_0(\Sigma)$ agrees
with the prescription defined by eq.~(\ref{dyn}) with $A_E = D$. $\Box$

\bigskip
\begin{center}
{\bf Acknowledgments}
\end{center}
We wish to thank Stefan Hollands and Itai Seggev for helpful
discussions. This research was initiated at the Yukawa Institute, and
R.M.W. wishes to thank the Yukawa Institute for its
hospitality. A.I. thanks the Enrico Fermi Institute for its
hospitality.  This research was supported in part by the Japan Society
for the Promotion of Science and by NSF grant PHY 00-90138 to the
University of Chicago.

\end{document}